\documentstyle[12pt,epsfig]{article}
\begin{document}

\title{Breakdown of the standard Perturbation Theory and Moving
  Boundary Approximation for ``Pulled'' Fronts} 
  
\author{Ute Ebert$^{1}$ and Wim van Saarloos$^2$\\
~\\
\small $^1$CWI,  Postbus 94079, 
  1090 GB Amsterdam, The Netherlands\\
\small $^2$Instituut--Lorentz, Universiteit Leiden, Postbus 9506,
  2300 RA Leiden,\\ 
\small The Netherlands }

\maketitle

\begin{abstract}
  The derivation of a Moving Boundary Approximation or of the response
  of a coherent structure like a front, vortex or pulse to external
  forces and noise, is generally valid under two conditions: the
  existence of a separation of time scales of the dynamics on the
  inner and outer scale and the existence and convergence of
  solvability type integrals. We point out that these
  conditions are not satisfied for pulled fronts propagating into an
  unstable state: their relaxation on the inner scale is power law
  like and in conjunction with this, solvability integrals diverge.
  The physical origin of this is traced to the fact that the important
  dynamics of pulled fronts occurs in the leading edge of the front
  rather than in the nonlinear internal front region itself. As recent
  work on the relaxation and stochastic behavior of pulled fronts
  suggests, when such fronts are coupled to other fields or to noise,
  the dynamical behavior is often qualitatively different from the
  standard case in which fronts between two (meta)stable states
  or pushed fronts propagating into an unstable state are considered.
\end{abstract}



\section{Introduction}

For a pattern in two or more dimensions that naturally can be divided
into domains and ``domain walls'' separating them, a much used
analytical approach is a moving boundary or effective interface
approximation
\cite{fife,karma,buckmaster,phfield1,phfield2,phfield3,phfield4,phfield5}.
This seems appropriate, when the width of the domain wall, front,
interface, or transition zone is much smaller than the typical length
scale of the pattern and when the dynamics of the pattern on long
space and time scales occurs through the motion of these interfaces.
The moving boundary approximation amounts to treating these fronts or
transition zones as a mathematically sharp interface or boundary. In
other words, their width is taken to be zero and their internal
degrees of freedom are eliminated. We shall henceforth use the word
boundary or interface to denote this zero width limit and use the word
front when we look at a scale where its internal structure can be
resolved.

Moving boundary approximations (MBA's) are ubiquitous in the theory of
pattern formation: they arise in most analytical approaches to late
stage coarsening \cite{gunton,bray}, in the analysis of interface
dynamics in dendritic growth and viscous fingering
\cite{langer,pelce,kkl,pomeau,brener,caroli,kassner}, step dynamics at
surfaces \cite{surfaces1,surfaces2,surfaces3}, thermal plumes
\cite{plumes,plumes2}, in chemical wave dynamics \cite{chemwaves},
combustion fronts \cite{buckmaster}, etc.

The main physical idea underlying the derivation of a MBA is that the
front itself can on large length and time scales be viewed as a
well-defined coherent structure which can be characterized by its
coordinates and a few effective parameters, such as its velocity or a
mobility coefficient. This idea plays a role for many coherent
structures, like vortices, or pulse-type solutions like sources,
sinks, solitons, etcetera
\cite{examples1,examples2,examples3,examples4,examples5}.  The
response of a coherent structure to an external driving force or noise
\cite{spanish,rocco} or the interaction between them can frequently be
derived by a perturbative expansion about the isolated coherent
structure solution. Often the effective parameters (a diffusion
coefficient, a mobility or an effective interaction force) 
can be derived from a solvability condition. A solvability condition
expresses that  a linear equation of the form  
$L\;\phi_1=g_1$, where the linear operator $L$ results from
linearizing about the isolated coherent structure solution, is
solvable provided $g_1$ is orthogonal to the kernel (null space) of
$L$. In other words, the requirement for such an equation to be
solvable is that
that $g_1$ is orthogonal to the left zero mode $\chi$ of $L$.

Although this is hardly ever mentioned explicitly, there are two
important implicit assumptions underlying such approximations, namely
$(a)$ that there is a separation of time scales between the motion of
the front as a whole and its internal dynamics, and $(b)$ that the
internal dynamics of the front is determined by the nonlinear front
region itself, so that the solvability type integrals are dominated by
the contributions from this finite region, and hence donot diverge.

The issue that we address in this paper is that while the above
conditions are satisfied for the familiar MBA for bistable fronts
and also for so-called pushed fronts,
they are not for so-called pulled fronts.  We will indeed discuss
several related properties of pulled fronts which bear on this: $(i)$
the divergence of the solvability integrals, with the concommittant
breakdown of a MBA or of the derivation of the response functions of
the front, like a diffusion or mobility coefficient; $(ii)$ the shift
of the dynamically dominant zone from the interior to the leading edge
of the front, that causes the solvability integrals to diverge;
$(iii)$ the fact that the stability spectrum of planar pulled front
solutions is gapless; $(iv)$ the recently discovered universal slow
power law relaxation of planar pulled fronts \cite{evs}. We will
initially focus our discussion on the derivation of a MBA, but as we
shall see our observations and conclusions apply equally well to
essentially any perturbative analysis of a pulled front.

The crucial feature of the standard moving boundary problem is that
the boundary conditions are {\em local} in space {\em and} time ---
e.g., the growth velocity of an interface is a function of the
instantaneous local temperature and curvature of the interface.
Usually, some of the boundary conditions are associated with
conservation laws, like the conservation of heat, and so they can
often be guessed from physical considerations.

If there is a separation of spatial scales, then such a MBA applies
only if there is also a separation of time scales between the internal
dynamics of the front and the dynamics of the outer bulk fields.
E.g., if the internal front modes relax on a time scale $\tau$, and
one considers a front of width $W$, propagation velocity $v$ and
typical curvature $\kappa$, then a MBA becomes appropriate in the
regime $\kappa W \ll 1$, $v\kappa \tau \ll 1$. Such a well-defined
relaxation time $\tau$ of a front on the inner scale actually exists
only if the relaxation is exponential in time. In this case, $\tau$ is
the inverse of the gap in the spectrum of the stability modes of the
planar front. Just like multiple scale and amplitude expansions
\cite{mult0,mult1,mult2} are based on projecting  all rapidly
decaying  gapped 
modes onto  the slow one (the center manifold), the MBA or
effective interface approximation can be thought of as projecting a
problem with fronts onto the slow interfacial dynamics.

However, if the stability spectrum of the planar front is gapless, the
internal modes of the front relax algebraically in time.  Thus there
is no characteristic time $\tau$ for the internal modes, no separation
of time scales and no standard MBA, no matter how thin the front
is. The 
internal dynamics of such a pulled front is actually slaved to the
evolution of its leading edge on the outer scale, which motivates the
term ``pulling''. Note that despite its different temporal behavior,
it is not at all visible on an instantaneous picture of a front,
whether it is bistable, pushed or pulled.

In a problem where the starting equations are
partial differential equations, the derivation of the MBA can often be
done analytically using by now standard methods. One should keep in
mind, however, that MBA's can be equally powerful in situations where
the approximation can not be derived cleanly by starting from a
partial differential equation and applying standard methods.  E.g., in
crystal growth the interfacial boundary conditions are determined on a
molecular scale, where for a rough interface the molecular processes
are so fast that after some coarse graining, we can describe the
interface for many purposes as a sharp interface whose response to
changes in temperature and concentration are instantaneous
\cite{vsalt}. Similar considerations apply to coarsening interfaces or
combustion fronts.

In the next section, we will first summarize the
necessary essentials of the stability and relaxation properties of
pulled fronts. Then, in section 3 we illustrate the issue by
following the standard derivation of a MBA for the type of coupled
equations that have in recent years been used in a phase-field type
formulation of solidification problems.  In section 4 we then discuss
the conditions under which such a type of analysis applies in more
detail, to identify the difficulties that arise when the front
dynamics on the inner scale is changed from the usual bistable or
pushed case to pulled. We then in section 5 generalize our findings
to equations with higher derivatives and to coupled equations, 
that create uniformly translating fronts. We show  that the
usual route of deriving solvability conditions does work in general 
for bistable and pushed fronts, but not for pulled fronts.

\section{Pulled fronts: Properties and statement of the problem}

When one considers a linearly unstable state, even a small
perturbation about this unstable state grows out and spreads.  We will
confine our analysis to fronts emerging from a localized initial
perturbation of the unstable state. One can calculate the asymptotic
linear spreading velocity $v^*$ of such a perturbation simply from the
linear dispersion relation $\omega(k)$ of the unstable modes according
to \cite{bers,ll}
\begin{equation}\label{lms}
  \left. \frac{{\rm d} \omega(k)}{{\rm d} k} \right|_{k^*}\!\!=\!v^*~,
    ~~~ \frac{ \mbox{Im}\, \omega(k^*)}{\mbox{Im}\, k^*} \!=\!v^*~.
\end{equation}
We furthermore will confine ourselves in this paper to fronts which
asymptotically are uniformly translating. For these, $\omega^*$ and
$k^*$ are purely imaginary, and we use the notation $k^*=i\lambda^*$.
If the above equation admits more than one solution, the one
corresponding to the largest value of $v^*$ is the relevant one. We
refer for the derivation of these results to our recent paper
\cite{evs}, which we will quote as paper I below.  Pulled fronts are
those for which the asymptotic spreading velocity $v_{as}$ of the
nonlinear front equals this linear spreading velocity $v^*$:
$v_{as}=v^*$ \cite{stokes,bj,vs2,paq,evs}. A number of model equations
for which fronts are pulled are discussed in paper I, but they also
arise in the analysis of more complicated situations like pearling
\cite{powers}, the Couette-Taylor instability \cite{ahlers},
Rayleigh-B\'enard convection \cite{fineberg}, the instabilities of
wakes of bluff bodies, leading e.g. to von Karman instabilities
\cite{provansal}, the emergence of global modes \cite{couairon},
liquid crystals \cite{liqcrys}, streamer discharge patterns
\cite{streamers}, the competion of domains in the Kupers-Lortz
instabilility \cite{tu}, the emergence of domains near structural
phase transitions \cite{salje}, polymer patterns \cite{polymers},
superconducting fronts \cite{dorsey}, error propagation
\cite{torcini}, deposition models \cite{krug}, step propagation
\cite{surfaces1}, chaotic fronts in the complex Ginzburg-Landau
equation \cite{nb,vsh,storm}, renormalization group analysis of
disorder models \cite{carpentier}, and the analysis of the Lyapunov
exponents in kinetic models \cite{vanzon}.

Fronts which propagate into an unstable state always are {\em pulled}
if all the nonlinearities suppress the growth. If not all of them do,
the asymptotic fronts speed $v_{as}$ may become larger than $v^*$:
$v_{as} = v^\dagger > v^*$. The relaxation of such ``{\em pushed}''
fronts \cite{stokes,paq,evs} is exponential with a characteristic
relaxation time $\tau $, that is finite \cite{vs2,evs}. As we will
discuss, for these the same perturbative schemes apply as for the
familiar bistable fronts, and likewise for these a standard type MBA
can be derived.

In paper I, we have shown that when a pulled front grows out of
sufficiently steep initial conditions (decaying into the unstable
state at least as $e^{-\lambda x}$ for $x \to \infty$ with some
$\lambda>\lambda^*$), then the velocity of a front obeys a {\em
  universal power law relaxation} given by
\begin{eqnarray} \label{predv}
  v(t)& \equiv & v^*+ \dot{X}(t)~, \\ \label{predv2} \dot{X}(t)& = &
  -\frac{3}{2 \lambda^* t} +\frac{3\sqrt \pi}{2\lambda^{*2} \sqrt{D}
    t^{3/2}} + { O}\left(\frac{1}{t^{2}}\right)~,
\end{eqnarray}
where
\begin{equation}
\label{D}
  D\!=\! \left. \frac{i {\rm d}^2 \omega(k)}{2 {\rm d} k^2}
\right|_{k^*}
\end{equation}
is real and positive for uniformly translating front solutions. For
the front profile, a similar power law relaxation holds, and the
extension of these results to one-dimensional pattern forming fronts
is given in \cite{storm}. The analysis reveals, that the power law
relaxation emerges from the dynamics of the foremost part of the
front where the dynamics is governed by the equations linearized
about the unstable state. The dynamics in the nonlinear region is
essentially slaved to this so-called leading edge. The
very slow $1/t$ power law relaxation of pulled fronts without
characteristic time scale obviously implies that the separation 
of time scales, which is necessary for a MBA to be applicable, is missing.

While from this perspective it is already intuively obvious that a
standard perturbation theory or MBA does not apply to pulled fronts,
the arguments underlying 
the separation of time scales are hardly ever discussed explicitly in
the literature on the derivation of a MBA. The purpose of this 
article therefore is to point out where the standard derivation breaks
down and how this emerges at a more formal level. In such an approach,
one generally encounters solvability type integrals of the form
\begin{equation}
  \int_{-\infty}^{\infty} d\xi \;e^{v\xi} \;\left(
  {{\partial\Phi_0}\over{\partial\xi}} \right)^2
\end{equation} 
or generalizations thereof --- see e.g.  (\ref{sol}), (\ref{solvab1})
or (\ref{solvagen}).  Here $\xi=x-vt$ is a frame moving with the front
with speed $v$, and $\Phi_0(\xi) $ is the associated planar front
solution.  The translation mode $\partial_\xi\Phi_0$ is a right
zero mode of the linear operator $L$ emerging from linearization about
the asymptotic front $\Phi_0$, and
$e^{v\xi}\;\partial_\xi\Phi_0$ is a left zero mode of this operator.
As we shall see, such solvability type integrals are well-defined and
finite for bistable and pushed fronts, but {\em diverge} for pulled 
fronts, since
the integrand does not converge for $\xi \to \infty$. In a way, the
solvability integral still correctly distributes its weight over the
dynamically important region, but for a pulled front, this region
becomes semi-infinite, and therefore the
integral diverges.  Our discussion also shows why introduction of an
ad-hoc cutoff in these integrals --- an approach that has sometimes
been considered in the literature --- does not necessarily cure the
problem.

\section{The derivation of a MBA from a phase field model} 

In this section, we first follow the standard derivation of a moving
boundary approximation (MBA) from a phase field model to highlight the
assumptions and approximations along the way. We then analyze
why and how the approximation breaks down for pulled fronts. 

As an example, we study the ``phase field model''
\begin{eqnarray}
\partial u / \partial t & = & \alpha\nabla^2 u + \partial \phi / \partial
t~,\label{phase1}
\\
\varepsilon \partial \phi / \partial t & = & \varepsilon^2 \nabla^2
\phi + f(\phi , u) ~, \label{phase2}
\\
\mbox{where }&&f(\phi , u) = \phi\;(1-\phi)\;(\mu-\lambda u+\phi)
~,~~~\lambda >0 ~.\label{f}
\end{eqnarray}
In the limit of zero front width $\epsilon\to0$, this model for
appropriate parameters $\alpha$, $\mu$ and $\lambda$ reduces to a
moving boundary approximation for a solidification front, where we can
think of $\phi$ as the order parameter field, while $u$ plays the role
of the temperature. $\phi$ then varies from the stationary
``liquid-like'' solution $\phi\approx 0$ in one domain to another
``solid-like'' solution $\phi\approx 1$ in the other domain.  Note
that in contrast to  \cite{fife}, $\partial
\phi / \partial t$ in (\ref{phase2}) has a coefficient $\varepsilon$,
not $\varepsilon^2$. This allows the front to have a velocity of order
unity, so the velocity is nonvanishing already in the lowest order
perturbation theory $O(\varepsilon^0)$. The $\partial \phi / \partial
t$ on the r.h.s.\ in (\ref{phase1}) models the generation of latent
heat in the interfacial zone where $\phi $ changes rapidly.

Other choices for $f(\phi ,u)$
can be found in the literature \cite{karma,phfield1,kupferman,phfield5}
but the form (\ref{f}) is most convenient for our present purpose. $f$
can be considered as the derivative of a ``free energy'' $F$,
\begin{eqnarray}
  f(\phi,u) &=& -\;{{\partial F (\phi ,u) }\over{\partial \phi}}~,\\
  F(\phi ,u) &=& -\;{{{(\mu-\lambda u) \;\phi^2}\over{2}}} -
  {{(1-\mu+\lambda u) \;\phi^3}\over{3}} + {{\phi^4}\over{4}}~.
\label{F}
\nonumber
\end{eqnarray}
Since $u$ varies on spatial and temporal scales of order unity, let us
treat it as a constant for a moment on the small length scale
$\varepsilon$ on which $\phi$ varies, and let us define
$\bar\mu=\mu-\lambda u$.  The connection with the phase field models
for solidification is closest in the range $-1 <\bar\mu <0$, when
the function $F$ (which is like a Ginzburg-Landau free energy
density), has two minima at $\phi=0$ and at $\phi = 1$.
When $\bar\mu=-1/2$, then
$F(0,u)=F(1,u)=0$ and the two ``phases'' $\phi=0$ and
$\phi=1$ are in equilibrium.  So if we choose the bare parameter
$\mu=-1/2$, then $u=0$ corresponds to the melting temperature, where
(\ref{phase2}) admits stationary front solutions with velocity $v=0$.
For $\mu=-1/2$ but $u$ nonzero, the minima of $f$ shift relative to
each other, and the order parameter front (\ref{phase2}) moves.  When
$u$ is positive, the liquid like minimum at $\phi=0$ is the absolute
minimum of $F$, and for $u$ negative the solid like minimum at
$\phi=1$ is the absolute one.  The front then will move such that
the state with the lowest free energy extends. For $\bar\mu>0$ the
state $\phi=0$ is linearly unstable; so we then deal with fronts
propagating into unstable states which are pushed for
$0<\bar\mu<1/2$ and pulled for $\bar\mu>1/2$ \cite{bj,vs2,evs}.
Though the interpretation of the model as a solidification model 
might be lost, we will illustrate the derivation of a MBA
as a function of $\bar\mu$ for this example, and we will find 
that the method breaks down at the transition from pushed to pulled 
fronts at $\bar\mu=1/2$.

Let us now trace the steps of the approximation in more detail.  The
field $u$ (\ref{phase1}) varies on a spatial scale of order unity, and
the field $\phi$ (\ref{phase2}) on a spatial scale of order
$\varepsilon\ll1$. A moving boundary approximation consists of first
matching an inner expansion of the problem on scale $\varepsilon$ to
an outer problem on scale 1, and then letting $\varepsilon\to0$ such
that an effective moving boundary problem on the outer scale results.
In the limit of $\varepsilon\to0$, the interface might have a
nonvanishing velocity and curvature on the outer length scale, so we
allow for $v=O(\varepsilon^0)$ and $\kappa=O(\varepsilon^0)$
\cite{noteKup}.

Let us for simplicity consider the problem in two spatial dimensions
$(x,y)$. On the outer scale, the fields are expanded in powers of
$\varepsilon$ as
\begin{eqnarray}
  u(x,y,t) &=& u_0(x,y,t)+\varepsilon\;u_1(x,y,t)+\ldots~,
\\
\phi(x,y,t) &=& \phi_0(x,y,t)+\varepsilon\;\phi_1(x,y,t)+\ldots~.
\end{eqnarray}
For a further analysis of these equations on the outer scale and their
matching to the inner scale, we refer to the literature. Here we focus
on the analysis of the $\phi$-front on the inner scale.
First a coordinate system moving with the front is introduced, where
$s$ measures the arc length of the interface in the tangential
direction, and $\xi$ the direction in which $\phi$ varies and
propagates. We put, e.g., $\xi=0$ at the place where $\phi=1/2$.
The coordinate $\xi$ in the direction normal to the front is   scaled
with a factor $\varepsilon$, since the front width will be of order
$\varepsilon$ in the limit $\varepsilon \to 0$. However, the
coordinate  $s$ is not scaled: along the front, the
variation is assumed to be simple on length scales of the order of
unity.  For the
inner expansion of the fields, one then writes 
\begin{eqnarray}
  u(x,y,t) &=& U_0(\xi,s,t)+\varepsilon\;U_1(\xi,s,t)+\ldots~,
\\
\phi(x,y,t) &=& \Phi_0(\xi,s,t)+\varepsilon\;\Phi_1(\xi,s,t)+\ldots~.
\label{defPhi}
\end{eqnarray}
The choice of coordinates can be illustrated when we consider a weakly
curved front which locally propagates with a velocity $v(s,t)$ in 
the $x$ direction, so that
\begin{equation}
  s= y ~~~,~~~\xi=\frac{x-X(s,t)}{\varepsilon}
  ~~~,~~~X(s,t)=x_0+\int^t {\rm d} t'~ v(s,t')~.
\end{equation}
In general, the front is curved and has a velocity $v$ and curvature
$\kappa$ which varies locally but on the outer time scale $t$ and
spatial scale $s$. They are therefore are expanded as
\begin{eqnarray}
  v(s,t) &=& v_0(s,t)+\varepsilon\;v_1(s,t)+\ldots~, \\ \kappa(s,t)
  &=& \kappa_0(s,t)+\varepsilon\;\kappa_1(s,t)+\ldots~.
\end{eqnarray}
The differential operators in (\ref{phase2}) then in the interior
coordinates $(\xi,s)$ have the $\varepsilon$ expansion
\begin{eqnarray}
  \varepsilon \left.\frac{\partial}{\partial t}\right|_{(x,y)}
  &=&\varepsilon \left.\frac{\partial}{\partial t}\right|_{(\xi,s)}
  -\Big[v_0+\varepsilon\;v_1+\ldots\Big]\;\frac{\partial}{\partial\xi}
  +O(\varepsilon^2)~,
\\
\varepsilon^2\;\nabla^2&=&\frac{\partial^2}{\partial\xi^2}
+\varepsilon\;\kappa_0\;\frac{\partial}{\partial\xi}+O(\varepsilon^2)~.
\end{eqnarray}
Inserting the expanded operators into (\ref{phase2}) and ordering in
powers of $\varepsilon$ yields in order $\varepsilon^0$
\begin{equation}
\label{Phi0}
\frac{\partial^2}{\partial\xi^2}\;\Phi_0
+v_0\;\frac{\partial}{\partial\xi}\;\Phi_0+f(\Phi_0,U_0)=0~,
\end{equation}
where $U_0$ is essentially constant on the inner scale $\xi$.  In
order $\varepsilon^1$ one finds
\begin{equation}
\label{Phi1}
L\;\Phi_1=-(\kappa_0+v_1)\;\frac{\partial}{\partial\xi}\;\Phi_0 +
\frac{\partial}{\partial t}\;\Phi_0 -\left.\frac{\partial
  f(\Phi_0,U)}{\partial U}\right|_{U_0}\;U_1
\end{equation}
with the linear operator
\begin{equation}
\label{L}
L \equiv \left( {{\partial^2 }\over{\partial \xi^2}} + v
{{\partial}\over{\partial\xi}} + \left. \frac{\partial
  f(\Phi,U_0)}{\partial \Phi}\right|_{\Phi_0} \right)~.
\end{equation}
Note that since $\Phi_0$ is a solution of the ordinary differential
equation (\ref{Phi0}), its time dependence occurs solely through the
variation  the $U$ field.

Equation (\ref{Phi1}) is an inhomogeneous linear differential equation
for the unknown field $\Phi_1$. If one has a left zero mode
$\chi(\xi)$ of $L$ such that $\chi L=0=L^\dag\chi$, where $L^\dag$
is the adjoint operator defined through partial integrations,
then (\ref{Phi1}) can be evaluated with a so-called ``solvability'' 
analysis by projection onto $\chi$:
\begin{equation}
\label{sol}
(\kappa_0+v_1)\;\int_{-\infty}^\infty d\xi\;\chi\;
\frac{\partial\Phi_0}{\partial\xi} +\int_{-\infty}^\infty
d\xi\;\chi\; \left.\frac{\partial f(\Phi_0,U)}{\partial
  U}\right|_{U_0}U_1= \int_{-\infty}^\infty
d\xi\;\chi\;\frac{\partial\Phi_0}{\partial t}~.
\end{equation}
There are clearly two important conditions for the identification
of (\ref{sol}) with the common solvability condition:
{\em If} the scalar products with $\chi$ exist, and 
{\em if} the temporal derivative $\partial_t\Phi_0$ of the 
zero order solution (\ref{Phi0}) can be neglected, 
{\em then} (\ref{sol}) expresses
the first order velocity correction $v_1$ as a function of the local
curvature $\kappa_0$, of the outer temperature field
$\partial_Uf\;U_1$ and of the zero order solution $\Phi_0$
(\ref{Phi0}).  It is exactly at these two points that the
analysis breaks down for pulled fronts. The violation of these
conditions always happens concomitantly, as they are
physically related.

Let us construct the left zero mode $\chi$ explicitly: It is well known,
that the right zero mode of $L$ is the mode of infinitesimal
translation $\partial_\xi\Phi_0$: $L\;\partial_\xi\Phi_0=0$.  Since
$L$ is nonhermitian, the left zero mode of $L$ is a right zero mode of
the adjoint $L^\dagger$ of $L$,
\begin{equation}
  L^\dagger \chi(\xi) = 0~,~~~~~~ L^\dagger \equiv \left( {{\partial^2
      }\over{\partial \xi^2}} - v {{\partial}\over{\partial\xi}} + \left. \frac{\partial
    f(\Phi,U_0)}{\partial \Phi}\right|_{\Phi_0} \right)~,
\end{equation}
and $\chi \neq \partial_\xi\Phi_0(\xi)$. However, the left zero mode
$\chi$ can be obtained by noting that the transformation
\begin{equation} 
  \phi = e^{-v\xi /2}\; \tilde{\phi}~, ~~~~~L \phi
  =\tilde{L}\tilde{\phi}~,
\end{equation}
with
\begin{equation}
\label{Ltil}
  \tilde{L} = e^{v\xi/2} \;L\; e^{-v\xi/2} = \left( {{\partial^2
      }\over{\partial \xi^2}} + \left. \frac{\partial
    f(\Phi,U_0)}{\partial \Phi}\right|_{\Phi_0}- {{v^2}\over{4}}
\right)~.
\end{equation}
turns the problem into a hermitian eigenvalue problem. As a result the
left zero eigenmode $\tilde{\chi}$ of $\tilde{L}$ is equal to the
right zero eigenmode $e^{v\xi/2}\partial_\xi\Phi_0$ of $\tilde{L}$.
Transforming back to $L$, this yields for the left zero mode
\begin{equation}
  \chi = e^{v\xi}\;\partial_\xi\Phi_0~,
\end{equation}
as can also be verified by substitution. If we may
ignore the term associated with the time derivative
$\partial_t\Phi_0$ and insert the expression for $\chi$ into
(\ref{sol}) we find
\begin{equation}
  v_1 = - \kappa_0 - \frac{\int_{-\infty}^{\infty} d \xi ~
    e^{v\xi} \; {{\partial \Phi_0 }\over{\partial \xi}}~ \left. {{\partial
        f(\Phi_0,U)}\over{\partial U}}\right|_{U_0} U_1 } {
    \int_{-\infty}^{\infty} d \xi ~ e^{v\xi} \;\left( {{\partial
        \Phi_0 }\over{\partial \xi}}\right)^2}~. \label{solvab1}
\end{equation}
If we furthermore ignore the term due to the coupling to the $u$ field, 
the expression $v_1= -\kappa_0$ is the familiar result of motion by mean
curvature first derived within the context of continuum models by
Allen and Cahn \cite{allencahn,bray}.

The structure of the solvability analysis is generic for the
perturbative expansion about a uniformly translating front.  Although
we have only considered the simplest type of model, and 
although refinements are possible \cite{karma}, Eq. (\ref{solvab1})
captures the basic structure of the expression that one obtains in
lowest order in a MBA: the relations between the velocity, curvature
and temperature field $u$ of the front, which play the role of
boundary conditions for the outer fields at the boundary in the zero
width limit $\varepsilon\to 0$, contain {\em solvability integrals} of
the form $\int d \xi\;e^{v\xi}\;(\partial_\xi \Phi_0)^2$. 
(Note that $\partial_Uf$ in (\ref{solvab1}) contains a
factor $\Phi_0$, that for $\xi\to\infty$ decays essentially like 
$\partial_\xi\Phi_0$.) Solvability
integrals of this type essentially arise in any type of perturbative
calculation, since they just express the solvability condition of the
linear perturbation problem $L \Phi_1(\xi) = g_1(\xi)$: the
inhomogeneous term $g_1(\xi)$ has to be 
orthogonal to the left zero mode $\chi$ of the linear operator $L$.

\section{Violation of the two conditions underlying  the MBA for
  pulled fronts}

We now discuss the conditions under which the MBA can be derived along
the lines sketched above in more detail.

Consider first the condition concerning the separation of time scales.
For fronts between two linearly stable states (with $-1<\bar\mu<0$
in $f$), there is such a separation
between the inner dynamics of the front and its displacement:
In this case, the stability spectrum of planar front modes has a gap
\cite{evs,vsalt}, and all internal eigenmodes decay as
$e^{-\omega_n t/\varepsilon}$ with eigenvalues 
$\omega_n\ge\omega_1=1/\tau=O(1)$.
Thus, in the limit $\varepsilon \to 0$, there is a clear separation 
of inner and outer time scales, and the adiabatic approximation 
(\ref{solvab1}) is justified on the outer time scale of order unity.
Moreover,
as discussed in paper I, for {\em pushed} fronts propagating into an
unstable state the stability spectrum is also gapped, and therefore
the separation of timescales necessary for the MBA to apply, does
hold. However, the stability spectrum of {\em pulled} fronts is
gapless, and as Eq.\ (\ref{predv}) of section II illustrates, pulled
fronts show indeed a power law convergence to their asymptotic speed
$v^*$. Clearly, then, the standard derivation of a
MBA does not apply to pulled fronts.

The same conclusion also emerges from the properties of the solvability
integrals themselves. For increasing $v$, the exponential factor
$e^{v\xi}$ enhances the value of the integrand for large positive
$\xi$, while suppressing the integrand for large negative $\xi$. We
therefore now turn to fronts propagating into an unstable state for
$\mu>0$ (for simplicity of notation, we use $u=0$), 
and investigate the behavior of the integrand for $\xi \to
\infty$. The large $\xi$ asymptotics of $\Phi_0(\xi)$ follows
directly from the {\em o.d.e.}\ (\ref{Phi0}) by noting that
$f'(\Phi_0(\infty))=f'(0)=\mu$, so that \cite{evs}
\begin{equation}
\Phi_0(\xi)  ~ \stackrel{\xi \gg 1}{\simeq}~ 
\left\{\begin{array}{ll}
A_1(v) \;e^{-\lambda_-\xi} + A_2(v) \;e^{-\lambda_+\xi}~, ~~~~ 
&v>v^*=2\sqrt{\mu}~,\label{phiv}\\
(\alpha \xi + \beta) \;e^{-\lambda^* \xi}~, ~~~~~~~~~~~~~~~~~~
 &v=v^*=2\sqrt{\mu}~,\label{phiv*}
\end{array}\right.
\end{equation}
where
\begin{eqnarray}
\lambda_\pm(v) &=& {{v}\over{2}} \pm {{1}\over{2}} \sqrt{v^2-4\mu}
 ~=~ {{v}\over{2}} \pm {{1}\over{2}}
 \sqrt{v^2-(v^*)^2} ~,\label{lambdapm}  ~~~\mbox{for }\mu>0\\ 
\lambda^* (v^*) & = & \lambda_\pm(v^*) = {{v^*}\over{2}}~.\label{lambda*}
\end{eqnarray}
The behavior of $\Phi_0$ for $v=v^*$ results from the fact that
precisely at the so-called pulled velocity $v^*$, the two roots
$\lambda_\pm$ coincide.

While for an arbitrary velocity $v>v^*$ the term $A_1(v)$ in
(\ref{phiv}) will be nonzero, so that the asymptotic behavior of
$\Phi_0$ is as $e^{-\lambda_-\xi}$, the pushed front solution --- if
it exists --- is precisely the solution with a well-defined value
$v=v^\dagger$ at which $A_1(v^\dagger)=0$. Note that for $\mu <0$,
we have $\lambda_-<0$, so that the relevant front solution in the range
$\mu<0$ has $A_1(v)=0$; thus the pushed front solution for $\mu>0$ is
precisely the analytic continuation of this front solution to the
regime $\mu>0$. If such a solution with $A_1(v^\dagger)=0$ exists, it
is the dynamically selected one from steep initial conditions \cite{evs}.
Moreover, these solutions decay for $\xi \gg 1$ as
$e^{-\lambda_+\xi}$, i.e., {\em faster} than $e^{-v^\dag\xi/2}$. As a
result, integrands in (\ref{solvab1}) like 
$e^{v^\dag\xi}\; (\partial_\xi\Phi_0)^2$ 
or $e^{v^\dag\xi} \;(\partial_\xi\Phi_0)\;g(\Phi_0,\xi)
\propto e^{v^\dag\xi} \;(\partial_\xi\Phi_0)\;\Phi_0$ for 
$\xi\to\infty$ are integrable, as
\begin{equation}
e^{v^\dag\xi} \left( {{\partial\Phi_0}\over{\partial\xi}}\right)^2
~~ \stackrel{\xi\gg1}{\simeq} ~~ e^{v^\dag\xi}\; e^{-2\lambda_+\xi} ~=~
e^{-\sqrt{(v^\dagger)^2-(v^*)^2}\;\;\xi} ~~
\stackrel{\xi\to\infty}{\longrightarrow}~  0~.\label{prodmodes}
\end{equation}

Thus, for a pushed front both criteria for a solvability analysis
of a perturbation theory are satisfied: the spectrum of the stability 
operator is gapped {\em and} the solvability integrals converge properly.

In passing, we note that the adjoint mode $\chi$ itself does not
decay to zero for large $\xi$ in the supercritical range $\mu>0$,
since
\begin{equation}
 \chi ~\stackrel{\xi\gg1}{\simeq}~ e^{v^\dagger \xi}\;
 {{\partial\Phi_0} \over {\partial\xi}} ~~\sim ~~ e^{\:\left(v^\dagger -
   \sqrt{(v^\dagger)^2-(v^*)^2}\;\right)\;\xi/2} ~~\stackrel{\xi \to
   \infty}{\longrightarrow}~~ \infty ~.
\end{equation}
For our perturbation theory this is no problem as long as
the inner product that defines the adjoint operator converges 
for $\xi\to \pm \infty$. Eq.\ (\ref{prodmodes}) shows that 
this is indeed the case.

While the solvability integrals converge properly for pushed fronts,
they do not for pulled fronts, as according to (\ref{lambda*}) 
\begin{equation}
e^{v^*\xi} \left( {{\partial\Phi_0}\over{\partial\xi}}\right)^2
~~ \stackrel{\xi\gg1}{\simeq} ~~\xi^2 \;e^{v^*\xi} \;e^{-2\lambda_*\xi}
~=~ \xi^2 ~~\stackrel{\xi\to\infty}{\longrightarrow}~ \infty ~.
\label{prodmodes*}
\end{equation}
As we already anticipated from the power law relaxation of pulled
fronts, standard perturbation theory used to derive a MBA does not
apply to pulled fronts. 

One could, of course, regularize the solvability integrals  by first
introducing a  cutoff $\xi_c$, and taking the cutoff to infinity as
the end of the calculation \cite{chen}. Whether such an approach
yields sensible 
results, depends on the situation under consideration. If, e.g., this
procedure is applied blindly to a solvability expression of the type
(\ref{solvab1}), one finds that the changes in the nonlinear terms of
the equation give no contribution --- in fact, since only the
divergent terms survive, this procedure amounts to calculating the
changes in $v^*$ in perturbation theory for changes in the parameters
in the linearized equation. Since $v^*$ can more easily be calculated
explicitly from Eq. (\ref{lms}) such a calculation has no particular
value. 

In fact, the divergence of the solvability integrals and the absence
of a characteristic time scale for the internal front dynamics
are deeply related. 
From (\ref{Ltil}) it is easily seen that the continuous
spectrum defined by $L\;\phi_\sigma = -\sigma \;\phi_\sigma$,
is bounded from below by $\sigma_0=(v^2-{v^*}^2)/4$.
For $\sigma (k)=\omega_0+k^2$, the eigenfunctions take the form
of Fourier modes $\phi_{\sigma (k)}\propto e^{\pm ik\xi}$
in the leading edge region $\xi\gg1$.
Hence for $v=v^*$, the gap $\sigma_0$ of the spectrum vanishes,
and all the eigenfunctions of $L$ are essentially plane waves
in the semi-infinite leading edge.
One finds furthermore \cite{evs}, that generic
perturbations of pulled planar fronts $\Phi_0$  are even  outside
the Hilbert space spanned by the eigenfunctions $\phi_\sigma$.
In this case, the long time dynamics cannot easily be understood
in terms of the eigenfunctions of $L$. One rather should directly 
study the linearized equation 
\begin{equation}
\varepsilon\partial_t\;\phi=\left[\varepsilon^2\;\nabla^2
+\left.\frac{\partial f(\phi,u)}{\partial\phi}
\right|_{\phi=0}\right]\;\phi+O(\phi^2)
\end{equation}
valid in the leading edge. In this formulation, the nonlinear
region of the front interior plays the role of a boundary condition
for the leading edge \cite{evs,storm}. As a result one finds predictions 
like (\ref{lms}) -- (\ref{D}).
Note finally, that the leading edge extends on the same outer length 
scale on which also $u$ varies. This demonstrates why it is not
possible to eliminate the dynamics of a pulled front
in a moving boundary approximation --- independent of how thin the front is.

\section{Generalization of the solvability analysis and of its break-down}

In the previous sections, we have traced the main steps in the
derivation of a MBA for two coupled equations that have been studied
as phase field models for solidification. In this case, the inner
equation for the order parameter reduces to the well-known nonlinear
diffusion equation studied first  by Fisher and Kolmogorov {\em et
al.} \cite{fisher,kolmogorov,aw}, and the nonhermitian linear
operator $L$ could be transformed to a hermitian operator $\tilde{L}$.
This allowed us to obtain the left zero mode $\chi$ of $L$
explicitly. When one considers higher order dynamical equations or
sets of coupled equations for the inner front region, it is usually
not possible to find the adjoint mode explicitly. Nevertheless, we
show in this section that the same conclusions hold more generally.

We consider a case where one has a vector $\vec{\phi}(x,t)$ of
dynamical fields, that in the long time limit can approach 
a planar uniformly translating front profile
$\vec{\Phi}_0(\xi)$ between the homogeneous stationary
states $\vec{\phi}^{\pm}=\vec{\Phi}_0(\pm\infty)$. The front
solution $\vec{\Phi}_0(\xi)$ with $\xi=x-vt$ obeys a set of {\em o.d.e.}'s, and
because of translation invariance $d\vec{\Phi}_0(\xi)/d\xi$ is again 
a zero mode of the linear matrix operator ${\bf L}$, obtained by
linearizing the {\em o.d.e.}'s about the front solution
$\vec{\Phi}_0(\xi)$:
\begin{equation} 
{\bf L} \left( {{d}\over{d\xi}}, {{d^2}\over{d\xi^2}},
{{d^3}\over{d\xi^3}}, \cdots ; \vec{\Phi}_0(\xi) \right) \cdot
{{d\vec{\Phi}_0 (\xi)}\over{d\xi}} =0~. \label{Lgen}
\end{equation}

If a front $\vec{\Phi}_0(\xi)$ is perturbed by external forces, 
other coherent structures or curvature effects, one generally
encounters equations like 
\begin{equation}
{\bf L}\cdot\vec{\phi}_1=\vec{g}_1
\end{equation}
in a perturbation expansion about $\vec{\Phi}_0$.
In our example above, $\vec{g}_1$ decayed essentially like 
$\vec\Phi_0$ and $d\vec\Phi_0/d\xi$ as $\xi\to\infty$,
and we only study such cases here. As is well known, such linear
equations are solvable provided the right hand side is orthogonal to
the kernel (null space) of $L$.
The existence of a left zero mode $\vec\chi$ of ${\bf L}$ therefore
generally leads to  the solvability condition
\begin{equation}
\label{solvagen}
\int_{-\infty}^\infty d\xi\;\vec{\chi}\cdot\vec{g}_1=0
~~~,~~~\vec{g}_1\stackrel{\xi\to\infty}{\sim}
{\bf Q}\cdot\frac{d\vec\Phi_0}{d\xi}~,
\end{equation}
(where the matrix {\bf Q} contains some slowly varying fields),
which relates parameters of the expansion as in (\ref{solvab1}).
So we now address the question of the existence of the left zero mode
$\vec\chi$ of ${\bf L}$, which is defined through 
${\bf L}^\dag\cdot\vec\chi=0$. In other words: 
$\vec{\chi}$ is the zero mode of the adjoint operator 
${\bf L}^\dagger$ obtained by partial integration,
\begin{equation}
\int_{-\infty}^{\infty} d\xi ~\vec{b}\cdot ({\bf L}\cdot \vec{a}) 
= \int_{-\infty}^\infty d\xi~ ({\bf L}^\dagger \cdot \vec{b} )\cdot 
\vec{a} ~.
\end{equation}
For this definition to hold, the integrals have to converge and
the boundary terms that arise from
performing the partial integrations all have to vanish.  This imposes
conditions on the allowed behavior of $\vec{b}$, given the asymptotic
behavior of $\vec{a}$: the product of these terms has to decay
sufficiently rapidly for $\xi \to \pm \infty$. 

In general, 
there is no particular simplifying relation between ${\bf L}$
and ${\bf L}^\dagger$; e.g., a term $f(\Phi_0)d/d\xi$ in
${\bf L}$ gives rise to a term $-\big(d f(\Phi_0)/d\xi\big)
- f(\Phi_0)d/d\xi$ in ${\bf L}^\dagger$. As a result, 
there is in general no simple relation between the left and right 
eigenmodes. However, since $\vec{\Phi}_0(\xi)$ approaches the constant
vectors $\vec\phi^\pm$ for $\xi\to \pm\infty$, the operators 
${\bf L}$ and ${\bf L}^\dag$ asymptotically are 
linear operators with constant coefficients, so that
\begin{eqnarray}
\lim_{\xi \to \pm\infty}& L^\dagger_{ij}& \left( {{d}\over{d\xi}}, 
{{d^2}\over{d\xi^2}},
{{d^3}\over{d\xi^3}}, \cdots ; \vec{\phi}^0_v(\xi) \right) = \nonumber
 L^\dagger_{ij} \left( {{d}\over{d\xi}}, {{d^2}\over{d\xi^2}},
{{d^3}\over{d\xi^3}}, \cdots ;\phi^\pm \right) = \\
&= &  L_{ji} \left(- {{d}\over{d\xi}}, {{d^2}\over{d\xi^2}},
- {{d^3}\over{d\xi^3}}, \cdots ;\phi^\pm \right) ~.\label{Ls}
\end{eqnarray}
Moreover, in this limit, the operator ${\bf L}$ is 
{\em exactly the same} as the one that one obtains from linearizing 
the set of {\em o.d.e.}'s for $\vec\Phi_0$ around the homogeneous 
stationary states $\vec{\phi}^\pm$. 
Therefore both $\vec\Phi_0$ and the right zero mode $d\vec\Phi_0/d\xi$
of ${\bf L}$ are asymptotically for $\xi\to\pm\infty$ just sums of simple
exponentials of the form 
\begin{equation} \label{phi1}
d\vec\Phi_0/d\xi~~  \stackrel{\xi\to \pm\infty}{\simeq} ~~
\sum_{n=1}^N \vec{a }_n^\pm \;e^{-\lambda_n^\pm \xi}~,
\end{equation} 
where the eigenvalues $\lambda_n^\pm$ are determined by
the characteristic polynomial of degree $N$
\begin{equation}
\det \;{\bf L} \Big(\;-\lambda_n^\pm \;, \;{\lambda_n^\pm\;}^2\;, 
\;-{\lambda_n^\pm\;}^3\;,\; \cdots \; ;\; \vec\phi^\pm\;\Big) = 0~. 
\label{roots}
\end{equation}
The asymptotic behavior of an adjoint zero mode $\vec{\chi}$ follows 
immediately from the symmetry relation (\ref{Ls}). If we write the
asymptotics for $\xi\to\pm\infty$ of $\vec{\chi}$ as 
\begin{equation}
\vec\chi~~\stackrel{\xi\to\pm\infty}{\simeq}~~
\sum_{n=1}^N \vec{b }_n^\pm \;e^{-\bar\lambda_n^\pm \xi}~,
\end{equation} 
then the eigenvalues $\bar{\lambda}_n^\pm$ 
are determined by the eigenvalue equation
\begin{equation}
\det \;{\bf L}^\dagger \Big(\;-\bar\lambda_n^\pm \;, 
\;{\bar\lambda_n^\pm\;}^2\;, 
\;-{\bar\lambda_n^\pm\;}^3\;,\; \cdots \; ;\; \vec\phi^\pm\;\Big)
= 0~, \label{roots2}
\end{equation}
which in view of (\ref{Ls}) and the fact that 
$\det{\bf L}=\det{\bf L}^\dagger$ immediately yields
\begin{equation}
\bar{\lambda}_n^\pm = - \lambda_n^\pm \label{lambdas}~.
\end{equation}

Let us now investigate the asymptotic behavior  of  products
$\vec{b}\cdot \vec{a}$ 
of left modes $\vec{b}$ and right modes $\vec{a}$, which is  required for the
existence and definition of the adjoint operator and modes.
Assume that the eigenvalues are ordered as
Re~$\lambda^\pm_{n+1}\ge$~Re~$\lambda^\pm_n$.
A pushed or bistable front is a discrete solution
with asymptotic behavior
\begin{equation}
\label{phidagger}\label{right}
\vec\Phi_0~~\simeq~~\left\{\begin{array}{cll}
\displaystyle\sum_{n=M+1}^N \vec{A }_n \;e^{-\lambda_n^+ \xi}&\simeq
\vec{A }_{M+1} \;e^{-\lambda_{M+1}^+ \xi}~~
&~\mbox{for }\xi\to\infty\\
\displaystyle\sum_{n=1}^M \vec{B }_n \;e^{-\lambda_n^- \xi}&\simeq
\vec{B }_{M} \;e^{-\lambda_{M}^- \xi}~~
&~\mbox{for }\xi\to-\infty
\end{array}\right.~,
\end{equation} 
that can be constructed from (\ref{phi1}) for a
particular value of $v=v^\dag$. 
The right zero mode $d\vec\Phi_0/d\xi$ obviously has the same 
asymptotic behavior.  In this expression, the eigenvalues $\lambda^-_1
, \cdots, \lambda^-_M $ for $\xi \to \infty$ are all the 
eigenvalues with negative real part so that the exponentials converge,
while on the right for 
$\xi \to \infty$ all $\lambda^+_{M+1}, \cdots,\lambda^+_{N}$ have
positive real parts. The existence of $M$ modes on the left and
$N-M+1$ modes on the right is a reflection of the fact that the
bistable or pushed front solution is an isolated (discrete) solution
\cite{evs}.

At this point, there is only one difference between bistable fronts and pushed 
fronts propagating into an unstable state: for the former,
Re~$ \lambda^+_{M} <0$ so that this mode is not present because it
corresponds to a diverging behavior, while for a pushed front
propagating into an unstable state, Re~$ \lambda^+_M >0$ but $A_M^+=0$
by definition \cite{evs}.  

A product of this right mode with 
a left mode converges to zero at $\pm \infty$, if the left zero mode
behaves asymptotically like 
\begin{equation}
\label{left}
\vec\chi~~\simeq~~\left\{\begin{array}{cll}
\displaystyle\sum_{n=1}^M \vec{C }_n \;e^{\lambda_n^+ \xi}&\simeq
\vec{C }_{M} \;e^{\lambda_{M}^+ \xi}~~
&~\mbox{for }\xi\to\infty\\
\displaystyle\sum_{n=M+1}^N \vec{D }_n \;e^{\lambda_n^- \xi}&\simeq
\vec{D }_{M+1} \;e^{\lambda_{M+1}^- \xi}~~
&~\mbox{for }\xi\to-\infty
\end{array}\right.~.
\end{equation} 
One easily verifies by counting the dimensions of stable and unstable
manifolds in the two asymptotic regions, that also $\chi$ belongs to a 
discrete spectrum, independent of the value of $M$, and that in
general the divergent term $\sim e^{\lambda_M^+\xi}$ is needed for
this mode to exist. 
Indeed, the textbook argument
\begin{eqnarray}
{\bf L}\cdot\vec\phi_m=\sigma_m\;\vec\phi_m~~~&,&~~~
{\bf L}^\dag\cdot\vec\chi_l=\sigma^\dag_l\;\vec\chi_l~,\label{cterm}
\\
\sigma^\dag_l\;\int \vec\chi_l\cdot\vec\phi_m
= \int ({\bf L}^\dag\cdot\vec\chi_l)\cdot\vec\phi_m
&=& \int \vec\chi_l\cdot({\bf L}\cdot\vec\phi_m)
= \sigma_m\;\int \vec\chi_l\cdot\vec\phi_m
\end{eqnarray}
shows that the eigenvalues $\sigma^\dag_l=\sigma_m$
equal each other, if the product of the eigenfunctions 
$\int \vec\chi_l\cdot\vec\phi_m$ is finite and likewise that
eigenfunctions with different eigenvalues are orthogonal. 
Application of simple ``counting arguments'' \cite{vsh,evs} for the
existence and multiplicity of solutions of {\em o.d.e.}'s shows that
(\ref{cterm}) implies that associated with the discrete right zero
mode of a pushed or bistable front solution, there is in general an
isolated (discrete) left zero mode of ${\bf L}^\dag$ with a nonzero
divergent term (\ref{cterm}).

This reasoning does not work for a pulled front, where the zero
mode of {\bf L} is part of a continuous spectrum with the same
asymptotic decay properties at $\xi\to\pm\infty$. 
The same counting argument as above now yields,
that in general no left zero mode of ${\bf L}^\dag$ exists.

This formal argument is supported by the observation,
that a solvability integral for a pushed front diverges
as the pushed velocity $v^\dag$ approaches the pulled velocity $v^*$
(\ref{lms}): Generally, the velocity $v$ will appear as a parameter 
in the characteristic polynomial (\ref{roots}). 
If we consider the $\lambda_n=\lambda_n^+$ as functions of $v$, 
then according to
the general scenario of front propagation into unstable states
\cite{evs} the pulled velocity is associated with a minimum of the
curve $v(\lambda_M) $ where $\lambda_M$ is the root of
(\ref{roots}) with the smallest positive real part. 
Hence for $v \stackrel{>}{\sim} v^*$ and uniformly translating
fronts with $\lambda_M$ and $\lambda_{M+1}$ real, we have
\begin{eqnarray}
\lambda_M(v) & = & \lambda^* - {{2}\over{v''}} \sqrt{v-v^*}+\cdots ~,
 \\
\lambda_{M+1}(v) & = & \lambda^* + {{2}\over{v''}} \sqrt{v-v^*}+\dots~,
 \end{eqnarray}
where
\begin{equation}
v'' = \left. {{d^2
      v(\lambda_M)}\over{d\lambda_M^2}}\right|_{\lambda^*}
\end{equation}
is the curvature of $v(\lambda_M)$ in the minimum that determines
$v^*$ and $\lambda^*$ (see \cite{evs}, section V.C.2). It hence is a
positive constant. 

In complete analogy with our earlier discussion in section 3, the
general scenario 
for front propagation into unstable states is that while the
asymptotic decay for $\xi \gg 1$ is as $e^{-\lambda_M\xi}$ for an
arbirary velocity $v$, a pushed front solution exists if for some
velocity $v^\dagger > v^*$, there is a front solution whose asymptotic 
large $\xi$ behavior is as $e^{-\lambda_{M+1}\xi}$ in agreement with
(\ref{phidagger}). 

If there is no such pushed front solution, then starting from
``steep'' initial conditions the selected front velocity is $v^*$; the 
asymptotic front profile with this velocity is then
\begin{equation}
\vec{\Phi}_0 (\xi)~ \stackrel{\xi \gg 1}{\simeq}~ (\vec{\alpha}\xi 
+ \vec{\beta})\; e^{-\lambda_M\xi}~~,~~\lambda_M=\lambda_{M+1} ~, 
\label{phiv*gen}
\end{equation}
in analogy with (\ref{phiv*}).

As we discussed above, for a pushed front, there is in general a
discrete left zero mode (\ref{left}) 
with asymptotic behavior $\chi\sim e^{\lambda_M\xi}$ for large $\xi$.
In spite of this divergence, the product of left and right modes converges as
\begin{eqnarray}
  \vec{\chi} \cdot {\bf Q} \cdot
  {{d\vec{\phi}_0}\over{d\xi}}~~ & \stackrel{\xi \gg
    1}{\simeq}&~~ e^{(\lambda_M-\lambda_{M+1})\xi}~ \vec{C}_M\cdot
  {\bf Q} \cdot \vec{A}_{M+1} ~,\nonumber \\ & \sim & e^{-(4/v'')
    \sqrt{v^\dagger - v^*}\;\xi}
  ~\stackrel{\xi\to\infty}{\longrightarrow}~ 0 ~~~(v
  \stackrel{>}{\sim} v^*)~,
\end{eqnarray}
and solvability conditions generally can be derived.

Just as we saw in the previous sections, the present analysis also shows
that as $v^\dagger$ approaches $v^*$ from above, the solvability
integrals converge  
less  and less fast until, at $v^*$, we have according to (\ref{phiv*gen}) 
\begin{equation}
 \vec{\chi} \cdot {\bf Q} \cdot
{{d\vec{\Phi}_0}\over{d\xi}}~~ \stackrel{\xi \gg 1}{\sim}~~
  \xi^2 
~~\stackrel{\xi\to\infty}{\longrightarrow}~~ \infty ~.
\end{equation}
in complete analogy with our earlier result (\ref{prodmodes*}) 
for the example discussed in section 3.

\section{Conclusions and outlook}

In contrast to ``bistable'' or pushed fronts, the dynamics of pulled
fronts is determined essentially in the leading edge. This was
recently shown to imply a general power law relaxation of pulled
fronts. In this paper, we have shown that this in turn entails that
pulled fronts lack the separation of time scales necessary for the
applicability of the usual MBA, and that solvability integrals diverge
when a front is pulled.

It is important to stress that one should not simply view this
negative result as a formal problem --- rather, one should take this
conclusion as a signal that the pattern dynamics involving the motion
of pulled fronts poses interesting new physical questions with
possibly surprising non-standard answers.

As a first simple illustration of this, consider the uncoupled F-KPP
equation (\ref{phase2}) in two dimensions with $\varepsilon=1$ and
$f=\phi-\phi^3$. If one starts with a radially symmetric steep initial
condition, e.g., $\phi(r,t=0)=\exp (-r^2)$, then this front will
spread out in a circularly symmetric way. According to (\ref{solvab1})
the curvature correction will then give a contribution $-1/r=
-1/(v^*t)=-1/(2t) $ to the velocity at large times. However, in
addition to that, there is a contribution $-3/(2t)$ of the same order
of magnitude from the power law relaxation (\ref{predv2}), as
$\lambda^*=1$ in this case. Thus, due to the combination of the power
law relaxation and the curvature correction, the front velocity $v^*$
will be approached asymptotically as $v(t)=v^*- 2/t$ \cite{derrida}!

In the example above of a circularly symmetric pattern without any
coupling to other fields, the relaxation and curvature effects can be
simply added up, but for a less trivial patterns whose shape is
changing in time, the proper description is far from obvious. 
Nontrivial patterns where pulled front propagation plays a
dominant role occur e.g.\ in streamer discharges \cite{streamers}.
Hence new analytical tools have to be developed for a moving boundary 
like description of these finger-like patterns. Work in progress 
\cite{arrayas} suggests that the limit of zero electron diffusion 
creating shock-like electron fronts is a valuable approximation
for negatively charged streamers.

A recent illustration of the fact that the nonexistence of
solvability integrals signals a transition to qualitatively different
dynamical behavior is given by the behavior of fronts in the presence
of multiplicative noise \cite{spanish,rocco}. Pushed fronts in
the presence of multiplicative noise show regular diffusive behavior
due to the noise being summed over the finite interior front region,
and their diffusion coefficient can be expressed in terms of solvability 
type integrals \cite{spanish}. In contrast, fully relaxed pulled fronts 
in an infinite system do not diffuse at all, and if a front with pulled
dynamics starts from a local (or ``sufficiently steep'' \cite{evs})
initial condition, it is subdiffusive \cite{rocco}: the root
mean square displacement of pulled
fronts increases with time as $t^{1/4}$, not as $t^{1/2}$. This
prediction was first suggested by using a time-dependent cutoff
$\xi_c(t)\sim \sqrt{t}$ in the solvability expression for the
diffusion coefficient that is valid for pushed fronts. The motivation
for this time-dependent cutoff comes from the relaxation analysis of
pulled fronts given in \cite{evs}. Hence, this example illustrates
both that using a cutoff in the solvability integrals sometimes {\em
  can} yield sensible results, and that the behavior of pulled fronts
can be qualitatively different from those of pushed fronts.

We finally note that these considerations also have implications for
numerical codes. In cases where a MBA applies in the limit in which
the front width is taken to zero, numerical codes with adaptive
gridsize refinement in the interior front region, where gradients are large,
are quite efficient. For pulled fronts, however, solutions with a too
coarse basic grid give inaccurate front velocities. For these, 
the refinement has to be done ahead of the
front, in the leading edge \cite{hundsdorfer}!

\section*{Acknowledgement}

The work of UE was supported by the Dutch research foundation
NWO and by the EU-TMR network ``Patterns, Noise and Chaos''.


\begin{thebibliography}{99}
\bibitem{noteKup} Kupferman et al.\ \cite{kupferman}
use the same $\varepsilon$ scaling in (\ref{phase2}), but even
allow for $\kappa=O(\varepsilon^{-1})$. However, this is not
consistent with the standard choice of the coordinate system 
on the inner scale \cite{fife} as also used by them and by us below.
\end{thebibliography}
\end{document}